\magnification=1200 \vsize=25truecm \hsize=16truecm \baselineskip=0.6truecm
\parindent=1truecm \nopagenumbers \font\scap=cmcsc10 \hfuzz=0.8truecm

\def\hypotilde#1#2{\vrule depth #1 pt width 0pt{\smash{{\mathop{#2}
\limits_{\displaystyle\tilde{}}}}}}

\null \bigskip  \centerline{\bf BILINEAR STRUCTURE AND SCHLESINGER TRANSFORMS}
\bigskip\centerline{\bf OF THE $q$-P$_{\rm III}$ AND $q$-P$_{\rm VI}$
EQUATIONS}
\vskip 2truecm
\centerline{{\scap M. Jimbo} and {\scap H. Sakai}}
\centerline{\sl Department of Mathematics}
\centerline{\sl Kyoto University }
\centerline{\sl Kyoto 606, Japan}
\bigskip
\centerline{\scap A. Ramani}
\centerline{\sl CPT, Ecole Polytechnique}
\centerline{\sl CNRS, UPR 14}
\centerline{\sl 91128 Palaiseau, France}
\bigskip
\centerline{\scap B. Grammaticos}
\centerline{\sl LPN, Universit\'e Paris VII}
\centerline{\sl Tour 24-14, 5${}^{\grave eme}$\'etage}
\centerline{\sl 75251 Paris, France}

\vskip 2truecm \noindent Abstract \smallskip
\noindent  We show that the recently derived ($q$-) discrete form of the
Painlev\'e VI equation
can be related to the discrete  P$_{\rm III}$, in particular if one uses
the full freedom in the
implementation of the singularity confinement criterion. This observation
is used here in order
to derive the bilinear forms and the Schlesinger transformations of both
$q$-P$_{\rm III}$ and
$q$-P$_{\rm VI}$.
\vfill\eject

\footline={\hfill\folio} \pageno=2

{\scap 1. Introduction.}
\medskip
\noindent In a recent paper [1], the first two authors obtained a first
order system of
$q$-difference equations the continuous limit of which was the sixth
Painlev\'e equation. This
completed the list of discrete forms of the Painlev\'e transcendents, since
a discrete P$_{\rm
VI}$ was the only one that was missing. The method used for obtaining this
mapping was
to consider a $q$-difference version of the monodromy preserving
deformation theory of
linear differential equations. (We recall the strong connection of the
continuous Painlev\'e equations [2] to the latter approach.) The
$q$-difference system that
was called $q$-P$_{\rm VI}$ is the following:
$$y \underline y ={a_3a_4(x-q^nb_1)(x-q^nb_2)\over(x-b_3)(x-b_4)}\eqno(1.1a)$$
$$\overline x  x ={b_3b_4(y-q^na_1)(y-q^na_2)\over(y-a_3)(y-a_4)}\eqno(1.1b)$$
where $x=x(n)$, $\overline x=x(n+1)$,
$\underline x=x(n-1)$ and $a_1,\dots,b_4$ are constants constrained by
$a_1a_2/a_3a_4=qb_1b_2/b_3b_4$. This system is integrable in the sense that
it results from the
compatibility of linear $q$-difference equations. Moreover, the singularity
confinement [3]
property of (1.1) was verified explicitly.

In [4], the remaining  authors, while missing the connection with  P$_{\rm
VI}$, identified
equation (1.1) as presumably integrable  in relation with  the discrete
P$_{\rm III}$, which was
called d-P$_{\rm III}$ (but should have been called $q$-P$_{\rm III}$ for
reasons that will
become obvious). Indeed, as was shown in [4], the first form of the
discrete P$_{\rm III}$ ever
obtained:
$$\overline x \underline x ={cd(x-a)(x-b)\over(x-c)(x-d)},\eqno(1.2)$$
with $c$, $d$ constant and $a$, $b$ proportional to $\lambda^n$,
is associated to a linear problem of $q$-difference type.

In section 2 we will show the precise relation between (1.1) and (1.2).
The method used for the derivation of  $q$-P$_{\rm VI}$ is the
``singularity confinement''
discrete integrability criterion. In section 3 we present the bilinearization of
both $q$-P$_{\rm III}$ and $q$-P$_{\rm VI}$ equations: the former is
obtained through the use of
four $\tau$-functions while for the latter 8 $\tau$-functions are
necessary. Section 4 is devoted
to the Schlesinger transformations of both equations.
\bigskip
{\scap 2. Derivation of the $q$-P$_{\rm VI}$ equation from singularity
confinement.}
\medskip
\noindent
We start with the $q$-P$_{\rm III}$ equation written in this general form
(where $a,b,c,d,g$
may {\sl a priori} all depend on $n$):
$$\overline x \underline x ={g(x-a)(x-b)\over(x-c)(x-d)}\eqno(2.1)$$
In order to obtain the precise integrable form of this equation we use the
singularity
confinement approach. This method is based on the observation that, {\sl
for an integrable
mapping}, the singularities that appear at some iteration, due to the
particular initial conditions
chosen, i.e. the `movable' singularities, disappear after some iterations:
they are ``confined''.
Based on this idea we can ask when a divergence appears in the iteration of
(2.1). This happens
whenever $x$ passes through one of the roots of the denominator. Let us
first consider the
case where $\underline x =\underline c$ while $\underline{\underline x}$ is
free.
We find thus that $x$ diverges and $\overline  x
\underline c=g$. Iterating further we would have found that the $x$'s
obtained do not in
general depend on the free quantity $\underline{\underline x}$. The only
way to restore
this dependence, i.e. to confine the singularity, at this stage, is to
balance the singularity of
$x$ by a singularity of the rhs of (2.1). Thus
$\overline  x$ must be equal to a root of the denominator. The assumption
$\overline  x=\overline  c$ is not acceptable because it corresponds to a
periodic singularity
and is thus in contradiction with the requirement that the singularity be
movable.
We are left with the choice
$\overline  x=\overline d$ and we have $\underline c \overline d=g$.
In a symmetric way, starting from $\underline x =\underline d$ we find
$\overline c \underline
d=g$. Thus the first condition for confinement is
$${\overline c/\overline d}={\underline c/\underline d}\eqno(2.2)$$
and $c/d$ is periodic of period two.

This is not, however, the only
source of divergence: $x$ may be equal to one of the roots of the
numerator. This would lead to
a diverging $\overline{\overline x}$ unless the divergence is balanced by
an appropriate
vanishing factor. Let us start with the case $\underline x =\underline a$.
This leads to $x=0$.
In order to obtain a finite $\overline{\overline x}$ we must have
$\overline x=\overline b$. (As previously, the confining condition
$\overline x=\overline a$ is
not acceptable because it corresponds to a non-movable singularity).  We
find that  we need
$\underline a \overline b  =gab/cd=\overline a \underline b$ where the last
equation is obtained
from the case $\underline x=\underline b$. Thus the second confinement
condition writes:
$${\overline a/\overline b}={\underline a/\underline b}\eqno(2.3)$$
So $a/b$ is also periodic of period two. In addition, using the
expression of $g$ above we also have
$$\underline a \overline b  cd=ab\overline c \underline d\eqno(2.4)$$
We remark at this stage that the two conditions (2.2-3) relate $n$'s of the
same parity. Thus, the
parameters with even and odd indices are only related by (2.4).

If one introduces the simplifying assumption that the ratios
$c/d$ and $a/b$ are not just periodic but strictly constant (as was done in
[4]), one finds simply
$q$-P$_{\rm III}$. A dependent variable transformation
$x(n)\to x(n)c(n)/C$ allows one to replace $c(n)$ by the constant $C$
(which we will denote
simply by $c$ from now on). Since $c/d$ is a constant, $d$ is now also a
constant. From (2.4) we
find that $a$ and $b$ are both proportional to $\lambda^n$ for some
$\lambda$.  However, this
simplification is unnecessary. Proceeding in full generality, we first use
the dependent variable
transformation (for even and odd indices separately) and impose $\overline
c=\underline c$ and thus
$\overline d=\underline d$. We now find that (2.4)  becomes $\underline a
\overline b  cd=ab\underline c \,\underline d$ and the same equation at
step $n+1$ is $ a
\overline{\overline b}
\underline c\,\underline d=\overline a\overline b cd$ i.e.
$${\overline a/\underline a}={\overline{\overline b}/ b}\eqno(2.5)$$
Combining  the latter with (2.3) implies that the ratio ${\overline
a/\underline a}={\overline
b/\underline b}$ must be a {\sl parity independent} constant. Calling this
ratio $q$  we find that
both $a$ and $b$ are proportional to $q^n$. We have in fact
$a_{2n}=q^{n}a_e$, $a_{2n+1}=q^{n}a_o$ (thereby breaking formally the
symmetry between even and
odd) and the analogous expression for $b$. Moreover, (2.4) now writes:
$$qa_eb_ec_od_o=a_ob_oc_ed_e \eqno(2.6)$$
If $q=\lambda^2$ and $a_o=\lambda a_e$, $b_o=\lambda b_e$, we recover the
$q$-P$_{\rm III}$ case.
As a last step we separate the even and odd $x$'s calling, for instance,
the odd ones $y$ with the
redefinition $x(2n)\to x(n)$, $x(2n+1)\to y(n)$, and similarly for the
$(a,b,c,d)$ which
at odd $n$'s  will now be called $(p,r,s,t)$. Because of the redefinition
of $n$, what was
previously called $\overline{\overline x}$ is now just $\overline x$. We
find {\sl two}
coupled equations for $x$ and $y$:
$$y \underline y ={st(x-a)(x-b)\over(x-c)(x-d)}\eqno(2.7a)$$
$$\overline x  x ={cd(y-p)(y-r)\over(y-s)(y-t)}\eqno(2.7b)$$
where $c,d,s,t$ are constants and $a,b,p,r$ are proportional to $q^n$. The
constraint (2.4) now
becomes:
$$prcd=qabst\eqno(2.8)$$

We recognize immediately in (2.7), with the appropriate renaming of the
coefficients, the system
(1.1) that defines the  $q$-P$_{\rm VI}$ equation.
Thus, $q$-P$_{\rm VI}$ is indeed contained in an embryonic state in
$q$-P$_{\rm III}$.

\bigskip
{\scap 3. Bilinear forms for $q$-P$_{\rm III}$ and $q$-P$_{\rm VI}$.}
\medskip
\noindent Let us start with the bilinearization of $q$-P$_{\rm III}$. We
expect $x$ to be given
by a rational expression involving the $\tau$-functions. The latter being
entire functions, the
divergences of $x$ will be associated to the vanishing of $\tau$-fuctions
which appear at the
denominator of $x$. In order to proceed, we need the precise singularity
structure of the
equation [5]. From the previous section, we know that a singularity appears
when $x$
passes through a root of the denominator. Then $x$ diverges at the next
step and the singularity
becomes confined at the subsequent step through the requirement that $x$
passes through
the other root. Two singularity patterns exist, $\{c,\infty,d\}$
and   $\{d,\infty,c\}$. The existence
of these two singularity patterns suggests the introduction of two
$\tau-$functions $F,G$:
$$x=c(1+{ \overline F \underline G \over FG})=d(1+{ \underline F \overline
G \over
FG}).\eqno(3.1)$$
The singularity patterns described above correspond to the  cases where
either $F$ or $G$ passes
through zero. Starting from this ansatz the bilinear form of $q$-P$_{\rm
III}$ was given in
[5]:
$$c\overline F \underline G -d\underline F \overline G +(c-d)FG=0\eqno(3.2a)$$
$${cd\over c-d}(c\overline {\overline F}\underline {\underline G}
-d\overline {\overline G}\underline {\underline F})+
(c-a)(d-b)FG+c(d-b)\overline F\underline G+d(c-a)\overline G\underline
F=0\eqno(3.2b)$$
The first equation comes just from equating the two expressions of $x$.
However, this is not the
simplest form. In fact, as we have seen, two `potential' singularities
exist when $x$ passes
through a root of the numerator. The patterns are $\{\underline a,0,
\overline b\}$ and
$\{\underline b,0,\overline a\}$ and we remark that no singularity actually
develops. This
suggests the introduction of two more, auxiliary
$\tau$-functions $J$, $K$ in the following manner:
$$x=c(1+{ \overline F \underline G \over FG})=d(1+{ \underline F \overline
G \over
FG})={JK\over FG}\eqno(3.3a)$$
$${1\over x}={1\over a}(1+{ \overline J\underline K\over JK})={1\over
b}(1+{ \underline J
\overline K \over JK})={FG\over JK}.\eqno(3.3b)$$
These equations in turn imply the $q$-P$_{\rm III}$ equation (1.2) for $x$.
By equating the
expressions of $x$ and $1/x$ we have four equations for the four
$\tau$-functions and thus
the bilinearization is trivially obtained:
$$c(FG+\overline F \underline G)=d(FG+\underline F \overline G)=JK\eqno(3.4a)$$
$${1\over a}(JK+\overline J\underline K)={1\over b}(JK+ \underline J
\overline K)=FG\eqno(3.4b)$$

In order to obtain the bilinear form for $q$-P$_{\rm VI}$ we start with the
same remark as for
the nonlinear variable: separating odd and even $x$'s we introduced a new
variable $y$.
Similarly, the
$\tau$-functions $F$ and $G$ (and also $J$, $K$ if we are talking about
(3.4)) are split into even
and odd. Let us keep the old names for the even ones, and for the odd
introduce the new
$\tau$-functions $M$, $N$, $P$, $Q$. We look now for $x$ and $y$:
$$x=c(1+{M \underline N \over FG})=d(1+{ \underline M N \over FG})={JK\over
FG}\eqno(3.5a)$$
$$y=s(1+{\overline F G \over MN})=t(1+{ F\overline G \over MN})={PQ\over
MN}\eqno(3.5b)$$
and the analogous expressions for $1/x$, $1/y$. However, the expressions
are slightly more
complicated if we leave full generality to $c$, $d$, $s$ and $t$. By rescaling
either $x$ or $y$ it is always possible to set
$$cd=st,\qquad qab=pr,\eqno(3.6)$$
which we will assume from now on.
Rewriting (3.2a) for both even and odd indices we find:
$$cM \underline N -d\underline M N+(c-d)FG=0$$
$$s\overline F G -t F\overline G +(s-t)MN=0\eqno(3.7a)$$
while (3.2b) becomes:
$${cd\over s-t}(s {\overline F} {\underline G}
-t {\overline G}{\underline F})+
(c-a)(d-b)FG+c(d-b)M \underline N+d(c-a)N\underline M=0$$
$${st\over c-d}(c {\overline M} {\underline N}
-d {\overline N}{\underline M})+
(s-p)(t-r)MN+s(t-r)\overline F G
+t(s-p)F \overline G=0\eqno(3.7b)$$
Similarly (3.4) can be transposed to the case of the 8 $\tau$-functions:
$$c(FG+M \underline N)=d(FG+\underline M N)=JK$$
$$s(MN+\overline F G)=t(MN+F\overline G)=PQ$$
$${1\over a}(JK+P\underline Q)={1\over b}(JK+ \underline P Q)=FG $$
$${1\over p}(PQ+\overline J  K)={1\over r}(PQ+ J \overline K)=MN\eqno(3.8)$$
where (3.6) has been assumed.
This completes the bilinearization of $q$-P$_{\rm VI}$.
\bigskip
{\scap 4. Miura and Schlesinger transforms for $q$-P$_{\rm III}$ and
$q$-P$_{\rm VI}$.}
\medskip
\noindent
In this section we shall derive the Schlesinger transformations for
$q$-P$_{\rm III}$ and
$q$-P$_{\rm VI}$. In the continuous case the Schlesinger are particular
auto-B\"acklund
transformations. As such they relate solutions of the same equation.
However, the Schlesinger
transformations relate solutions corresponding to the same monodromy data
except for {\sl
integer} differences in the monodromy exponents [2]. In the discrete case
the relation of the
Schlesinger transformations to monodromy exponents is not always clear.
However, we can use an
analogy with the continuous case. If one uses the proper parametrisation of
the equation the
Schlesinger transformations can be shown to be associated to simple changes
of the parameters.
The discrete case can be analysed in the same spirit. By using the proper
parametrisation, one
can identify, among the auto-B\"acklund transformations, those which
correspond to simple
changes  of the parameters and which can thus be dubbed Schlesinger's.
As in the previous section, we shall start with $q$-P$_{\rm III}$ and then
generalize
the results to $q$-P$_{\rm VI}$ that is viewed as an asymmetric $q$-P$_{\rm
III}$.

In order to find the Miura transform of $q$-P$_{\rm III}$, we use the
standard `trick'
introduced in [6]. Starting from the variable $x$ expressed in terms of
{\sl two} $\tau$-functions
$F$ and $G$, we introduce a new variable $u$ which can be expressed in
terms of only {\sl one}
$\tau$-function, say $G$. We remark that $(x-c)$ is proportional to
$\overline F/F$ while $(x-d)$
contains $\underline F/F$. Upshifting the last object and multiplying by
the first allows to get
rid of $F$ entirely. We thus find a first relation between
$x$ and $u$:
$$u=(x-c)(\overline x-d)\eqno(4.1)$$
where $u$ is given by $u=cd\underline G \overline{\overline G}/G\overline
G$. This is  the first half of the Miura. Eliminating $F$ between the two
bilinear equations (3.2)
leads to a (hexalinear) equation for $G$ which can also be expressed in
$u$. The latter
would be the `modified' equation of $q$-P$_{\rm III}$ (`modified' in the
sense of the relation
that exists between the Korteweg-de Vries equation and the modified
Korteweg-de Vries equation).
However, it is simpler to obtain the second half of the Miura and proceed
from there. As we have
shown in [6], the complement of the Miura involves a rational expression
which must be homographic
in both $u$ and $\underline u$. We find readily:
$$x= {u \underline u/cd- u-\underline u+cd-ab\over - u/d-\underline
u/c+c+d-a-b}\eqno(4.2)$$

Eliminating $x$ and $\overline x$ between (4.1), (4.2) and its upshift
leads to an equation for
$u,\underline u,\overline u$. This equation is, after a change of
variables, a discrete form of
$q$-P$_{\rm V}$ equation (although not all the parameters of a $q$-P$_{\rm
V}$ are present).
Introducing $U=u-cd$, we find:
$$(U\overline U-\lambda^2abcd)(U\underline
U-abcd)={cd(U+bd)(U+\lambda ac)(U+ad)(U+\lambda bc)\over U+cd}\eqno (4.3)$$
In order to define a different Miura we can introduce the quantity:
$$w=(1/x-1/a)(1/\overline x-1/\overline b)\eqno(4.4)$$
(we recall that $\overline b=b(n+1)=\lambda b(n)$) which depends only on
the $\tau$-function $K$
but not $J$. The second half of the Miura, analogous to (4.2), is:
$$x={-a\underline w/\lambda-bw\lambda+1/a+1/b-1/c-1/d\over abw\underline
w+1/ab-1/cd-w\lambda-\underline w/\lambda}\eqno(4.5)$$
Again eliminating $x$ leads to an equation for $w$. Introducing
$W=w-1/a\overline b$, we find
for this equation:
$$(W\overline W-{1\over \lambda^2 abcd})(W\underline W-{1\over abcd}) =
{(W+1/\lambda bd)(W+1/ac)(W+1/\lambda bc)(W+1/ad)\over \lambda abW+1}\eqno
(4.6)$$
The quantity $\tilde W=U/\lambda abcd$ satisfies an equation, obtained from
(4.2), namely:
$$(\tilde W\tilde{\overline W}-{1\over \lambda^2 abcd})(\tilde W\underline
{\tilde
W}-{1\over abcd}) = {(\tilde W+1/\lambda ac)(\tilde W+1/bd)(\tilde W+1/\lambda
bc)(\tilde W+1/ad)\over \lambda ab\tilde W+1}\eqno (4.7)$$
Equation (4.7) has the same form as (4.6) provided one introduces the
parameters
$$\tilde a=a\sqrt\lambda,\ \tilde b=b/\sqrt\lambda,\ \tilde
c=c\sqrt\lambda,\ \tilde
d=d/\sqrt\lambda\eqno(4.8)$$
We define
$$\tilde w=\tilde W+1/\lambda\tilde a\tilde b=u/\lambda abcd\eqno(4.9)$$ and
$$\tilde x={-\tilde a\underline {\tilde w}/\lambda-\tilde b\tilde
w\lambda+1/\tilde
a+1/\tilde b-1/\tilde c-1/\tilde d\over \tilde a\tilde b\tilde w\underline
{\tilde w}+1/\tilde a\tilde b-1/\tilde c\tilde d-\tilde w\lambda-\underline
{\tilde w}/\lambda}\eqno(4.10)$$
Given this definition of $\tilde x$ and since $\tilde W$ satisfies (4.7) it
follows that
$$\tilde w=(1/\tilde x-1/\tilde a)(1/\tilde {\overline x}-1/\tilde
{\overline b})\eqno(4.11)$$
and therefore $\tilde x$ satisfies $q$-P$_{\rm III}$ with parameters
$\tilde a, \tilde b, \tilde
c, \tilde d$. The transformation from $x$ to $\tilde x$ through (4.1),
(4.8), (4.9) and (4.10)
defines an auto-B\"acklund transformation for $q$-P$_{\rm III}$. In this
case this is
indeed a Schlesinger transformation which we denote by $S_c^a$. (The
convention used here
is to give explicitly the parameters associated to $x$, rather than
$\overline x$, in (4.1) and
(4.4).)  The inverse transformation  $(S_c^a)^{-1}$ can be obtained by
defining $w$ through
(4.4), $\hypotilde{0}{u}=w\lambda abcd$ and finally $\hypotilde{0}{x}$
through the analog of
(4.2).

In a similar way we can introduce the transformations $S_c^b$,
$S_d^a=(S_c^b)^{-1}$ and
$S_d^b=(S_c^a)^{-1}$. They correspond to multiplying the two  parameters
which appear explicitly
by $\sqrt\lambda$ while dividing the two others by the same quantity. These
are the most
elementary Schlesinger transformations. Using them we can construct further
Schlesinger
transformations that act separately on $\{a,b\}$ or $\{c,d\}$. For instance
the product
$S_c^aS_d^a$ corresponds to  $a\to a\lambda,\quad b\to b/\lambda,\quad c\to
c, \quad d\to d $.

We turn now to $q$-P$_{\rm VI}$ and its Schlesinger transforms. Given its
relation to $q$-P$_{\rm
III}$ it is straightforward to find a Miura transformation analogous to
(4.1). Note however that
we need a two-component Miura:
$$u=(x-c)(y-t),\qquad v=(y-s)(\overline x-d)\eqno(4.12)$$
and its second half:
$$x= {u\underline v-sdu-ct\underline v+st(cd-ab)\over -su-t\underline
v+st(c+d-a-b)}\eqno(4.13a)$$
$$y= {uv-sdu-ctv+cd(st-pr)\over -du-c v+cd(s+t-p-r)}\eqno(4.13b)$$
Eliminating $x$ and $y$ between (4.12) and (4.13) leads to an equation for
$u,v,\underline v$ while eliminating $\overline x$ and $y$ from (4.12),
(4.13b) and the upshift
of (4.13a) leads to an equation for $u,v,\overline u$. We introduce, just
as in the case of
$q$-P$_{\rm III}$, the new variables $U=u-ct$, $V=v-sd$ and we obtain:
$$(U\underline V-abst)(UV-prcd)={sd(U+cr)(U+bt)(U+cp)(U+at)\over
U+ct}\eqno(4.14a)$$
$$(VU-prcd)(V\overline U-q^2abst)={ct(V+qsb)(V+rd)(V+qas)(V+pd)\over
V+sd}\eqno(4.14b)$$
In parallel to (4.4) we introduce the Miura:
$$w=(1/x-1/a)(1/y-1/r),\qquad z=(1/y-1/p)(1/\overline x-1/\overline
b)\eqno(4.15)$$
where now $\overline b=qb$, the second half being:
$$x= {-qrw-p\underline z+q(1/a+1/b-1/c-1/d)\over prw\underline
z-qrw/b-p\underline
z/a+q(1/ab-1/cd)}$$
$$y= {-aw-qbz+(1/p+1/r-1/s-1/t)\over qabwz-aw/p-qbz/r+(1/pr-1/st)}\eqno(4.16)$$
The quantities $W=w-1/ar$ and $Z=z-1/qpb$ satisfy the system:
$$(W\underline Z-q/prcd)(WZ-1/qabst)={(W+1/at)(W+1/dr)(W+1/as)(W+1/cr)\over
pb(W+1/ar)}
\eqno(4.17a)$$
$$(ZW-1/qabst)(Z\overline
W-1/qprcd)={(Z+1/pd)(Z+1/qbt)(Z+1/pc)(Z+1/qbs)\over qar(Z+1/qpb)}
\eqno(4.17b)$$
In analogy to the $q$-P$_{\rm III}$ case we introduce $\tilde W=U/\lambda
abst$ and $\tilde
Z=V/\lambda prcd$, (recall that $q=\lambda^2$), and obtain:
$$(\tilde W\underline {\tilde Z}-q/prcd)({\tilde W}{\tilde Z}-1/qabst)=
{d({\tilde W}+\lambda/dr)({\tilde W}+1/\lambda bs)({\tilde W}+\lambda/pd)
({\tilde W}+1/\lambda
as)\over \lambda abt({\tilde W}+c/\lambda abs)}\eqno(4.18a)$$

$$({\tilde Z}{\tilde W}-1/qabst)({\tilde Z}\tilde {\overline
W}-1/qprcd)={t({\tilde
Z}+1/\lambda at)({\tilde Z}+1/\lambda pc) ({\tilde Z}+1/\lambda bt)({\tilde
Z}+1/\lambda rc)\over
\lambda prd({\tilde Z}+s/\lambda rpc)}\eqno(4.18b)$$
This system has the same form as (4.17) provided one introduces the parameters:
$$\tilde a=\sqrt{aps\over c},\
\tilde b={1\over \lambda}\sqrt{brc\over s},\
\tilde c=\sqrt{pcs\over a},\
\tilde d=\sqrt{adt\over p}\eqno(4.19)$$
$$\tilde p=\lambda\sqrt{apc\over s},\
\tilde r=\sqrt{brs\over c},\
\tilde s=\lambda \sqrt{acs\over p},\
\tilde t={1\over \lambda}\sqrt{pdt\over a}$$
We remind the reader that in section 3 we fixed the relative scaling of $x$
and $y$ so that
$cd=st$ and $qab=pr$ (equation 3.6). It is easy to check that these
relations are satisfied by
the tilded parameters above.
We define
$$\tilde w=\tilde W+1/\tilde a\tilde r=u/\lambda abst,\qquad\tilde z=\tilde
Z+1/q\tilde
p\tilde b=v/\lambda prcd\eqno(4.20)$$ and
$\tilde x,\tilde y$ by the equivalent of (4.16) where $w,z$ and all the
parameters are tilded.
Again, since $\tilde W,\tilde Z$ satisfy (4.18) we have
$$\tilde w=(1/\tilde x-1/\tilde a)(1/\tilde y-1/\tilde r),\qquad \tilde
z=(1/\tilde
y-1/\tilde p)(1/\tilde {\overline x}-1/\tilde {\overline b})\eqno(4.22)$$
and $\tilde x,\tilde y$ satisfy
$q$-P$_{\rm VI}$ with `tilded' parameters. The transformation from $x,y$ to
$\tilde x,\tilde y$ is
an auto-B\"acklund transformation $B_{cs}^{ap}$ for $q$-P$_{\rm VI}$ that
is {\sl not} a
Schlesinger. (The convention again is that $a$ are associated to $x$ in
$w$, $p$ to $y$ in $z$,
$c$ is associated to $x$ in $u$ and $s$ to $y$ in $v$). Starting from that
auto-B\"acklund
$B_{cs}^{ap}$ it is straightforward to obtain a Schlesinger transformation.
In fact
$(B_{cs}^{ap})^2$ is such a transformation where
$$a\to a \lambda,\ b\to b/ \lambda,\ c\to c \lambda,\ d\to d/ \lambda,\
p\to p \lambda,\ r\to r/ \lambda,\ s\to s \lambda,\ t\to t/ \lambda\eqno(4.23)$$
However this is not the simplest Schlesinger transformation one can
construct starting from $B$.
In fact, through the product of auto-B\"acklund transformations
$B_{ct}^{ar}B_{cs}^{ap}$ we
obtain the parameters:
$$\hat a=\sqrt{abd\over c},\
\hat b=\sqrt{abc\over d},\
\hat c=\sqrt{bcd\over a},\
\hat d=\sqrt{acd\over b}\eqno(4.24)$$
$$\hat p={1\over \lambda}\sqrt{prt\over s},\
\hat r=\lambda \sqrt{prs\over t},\
\hat s={1\over \lambda}\sqrt{rst\over p},\
\hat t={\lambda}\sqrt{pst\over r}$$
Next we perform the transformation, $X=\sqrt{abcd}/\hat x$,
$Y=\sqrt{prst}/\hat y$ and obtain
$q$-P$_{\rm VI}$ for the variables $X,Y$. The resulting transformation is a
Schlesinger
corresponding to parameters:
$$a\to a,\ b\to b,\ c\to c,\ d\to d,\
p\to p \lambda,\ r\to r/ \lambda,\ s\to s \lambda,\ t\to t/ \lambda\eqno(4.25)$$
Further Schlesinger transformations can be obtained through combinations of
the appropriate
$B$'s.
\bigskip {\scap 5. Conclusion}
\medskip
\noindent In this paper we have undertaken a study of the discrete P$_{\rm
VI}$ ($q$-P$_{\rm
VI}$) equation that was derived in [1]. We have shown that this equation
can be viewed as a kind
of `asymmetric' $q$-P$_{\rm III}$. Indeed, its form is the most general one
compatible with the
singularity confienement requirement when applied to a $q$-P$_{\rm
III}$-type ansatz. Using
this analogy we were able to obtain, in close parallel, results for the
$q$-P$_{\rm III}$ and
$q$-P$_{\rm VI}$. in particular we gave a simple bilinearization for both
equations, involving
four $\tau$-functions in the $q$-P$_{\rm III}$ and eight $\tau$-functions
in the $q$-P$_{\rm VI}$
case. By deriving the appropriate Miura transformations for both equations
we obtained their
Schlesinger transformations.

Several open questions remain at this stage. For example, the structure of
the space of
particular solutions of $q$-P$_{\rm VI}$ has only been touched upon in [1].
Rational solutions
have not been considered at all. The coalescence chain of $q$-P$_{\rm VI}$
which is expected to
contain new discrete Painlev\'e equations has not been worked out yet. We
hope to return to
these questions in future publications.
\bigskip {\scap Acknowledgements}.
\medskip
\noindent The authors acknowledge interesting discussions with J. Satsuma
and Y. Ohta.

\bigskip {\scap References}.
\medskip
\item{[1]}      M. Jimbo and H. Sakai, {\sl A $q$-analog of the sixth
Painlev\'e equation}, preprint
Kyoto-Math 95-16.
\item{[2]} M. Jimbo, T. Miwa, Physica D2 (1981) 407.
\item{[3]} B. Grammaticos, A. Ramani and V. Papageorgiou, Phys. Rev. Lett.
67 (1991) 1825.
\item{[4]} A. Ramani, B. Grammaticos and J. Hietarinta, Phys. Rev. Lett. 67
(1991) 1829.
\item{[5]} A. Ramani, B. Grammaticos and J. Satsuma, Jour. Phys. A 28
(1995) 4655.
\item{[6]} A. Ramani and B. Grammaticos, Jour. Phys. A 25 (1992) L633.

\end